\documentclass[aps,prb,onecolumn,showpacs,preprint,amsmath,amssymb,groupaddress]{revtex4}

\usepackage{graphicx}
\usepackage{dcolumn}
\usepackage{bm}
\usepackage{subfigure}

\begin{document}

\title{Magnetic Graphene Nanohole Superlattices}

\author{Decai Yu,  Elizabeth M. Lupton, Miao Liu, Wei Liu and Feng Liu}
\email{fliu@eng.utah.edu} \affiliation{Department of Materials
Science and Engineering, University of Utah, Salt Lake City, UT
84112}

\begin{abstract}
We investigate the magnetic properties of nano-holes (NHs) patterned
in graphene using first principles calculations. We show that
superlattices consisting of a periodic array of NHs form a new
family of 2D crystalline "bulk" magnets whose collective magnetic
behavior is governed by inter-NH spin-spin interaction. They exhibit
long-range magnetic order well above room temperature. Furthermore,
magnetic semiconductors can be made by doping magnetic NHs into
semiconducting NH superlattices. Our findings offer a new material
system for fundamental studies of spin-spin interaction and magnetic
ordering in low dimensions, and open up the exciting opportunities
of making engineered magnetic materials for storage media and
spintronics applications.
\end{abstract}


\maketitle

Magnetic materials have a wide range of applications, such as being
used for storage media. Magnetism is commonly associated with
elements containing localized \emph{d} or \emph{f} electrons, i.e.
the itinerant ferromagnetism\cite{Slater,FLiu}. In contrast, the
elements containing diffuse \emph{sp} electrons are
\emph{intrinsically} non-magnetic, but magnetism can be induced in
\emph{sp}-element materials extrinsically by defects and impurities.
There have been continuing efforts in searching for new magnetic
materials, and much recent interest has been devoted to magnetism of
carbon-based\cite{Shibayama,Esquinazi,Lehtinen,HLee,Coey},
especially graphene-based
structures\cite{Kusakabe,YWSon,Nomura,Pisani,Brey,Novoselov,Jiang,Yazyev,Fernandez,Wang,Huang}
such as graphene nanoribbons\cite{Kusakabe,YWSon,Pisani,Huang}
 and nanoflakes\cite{Fernandez,Wang}. Here, we predict a new class of
graphene-based magnetic nanostructures, the superlattices of
graphene nanoholes (GNHs), using first-principles calculations.

Graphene nanoribbons\cite{Kusakabe,YWSon,Pisani,Huang} and
nanoflakes\cite{Fernandez,Wang} with zigzag edges have been shown
to exhibit magnetism. Their magnetization is originated from the
localized edge states that give rise to a high density of states
at the Fermi level rendering a spin-polarization instability\cite{Slater}. Then, for the same reason, if nanoholes
(NHs) are made inside a graphene sheet with zigzag edges they may
also exhibit magnetism. Furthermore, by making an array of NHs, we
may expect collective "bulk" magnetism because \emph{inter}-NH
spin-spin interactions are introduced in addition to the intra-NH
spin coupling. This allows us to go beyond the current scope
limited to the spins within a single nanoribbon or nanoflake. In
effect, superlattices consisting of a periodic array of NH spins
form a family of nanostructured magnetic 2D crystals with the NH
acting like a "super" magnetic atom.

We have investigated magnetic properties of GNHs, using
first-principles pseudopotential plane-wave calculations within
the spin-polarized generalized gradient approximation
\cite{Kresse}. We used a rhombus supercell in the graphene plane
with the cell size ranging from 14$\times$14{\AA} to
41$\times$41{\AA} and a vacuum layer of $\sim$10{\AA}. We used a
2$\times$2$\times$1 k-point mesh for Brillouin zone sampling and a
plane wave cutoff of 22.1 Rd. The systems contain up to a maximum
of 530 atoms. The dangling bonds on the edge atoms are saturated
with hydrogen. The system is relaxed until the force on each atom
is minimized to less than 0.01 eV/{\AA}.

\begin{figure}[h]
\includegraphics[width=6in]{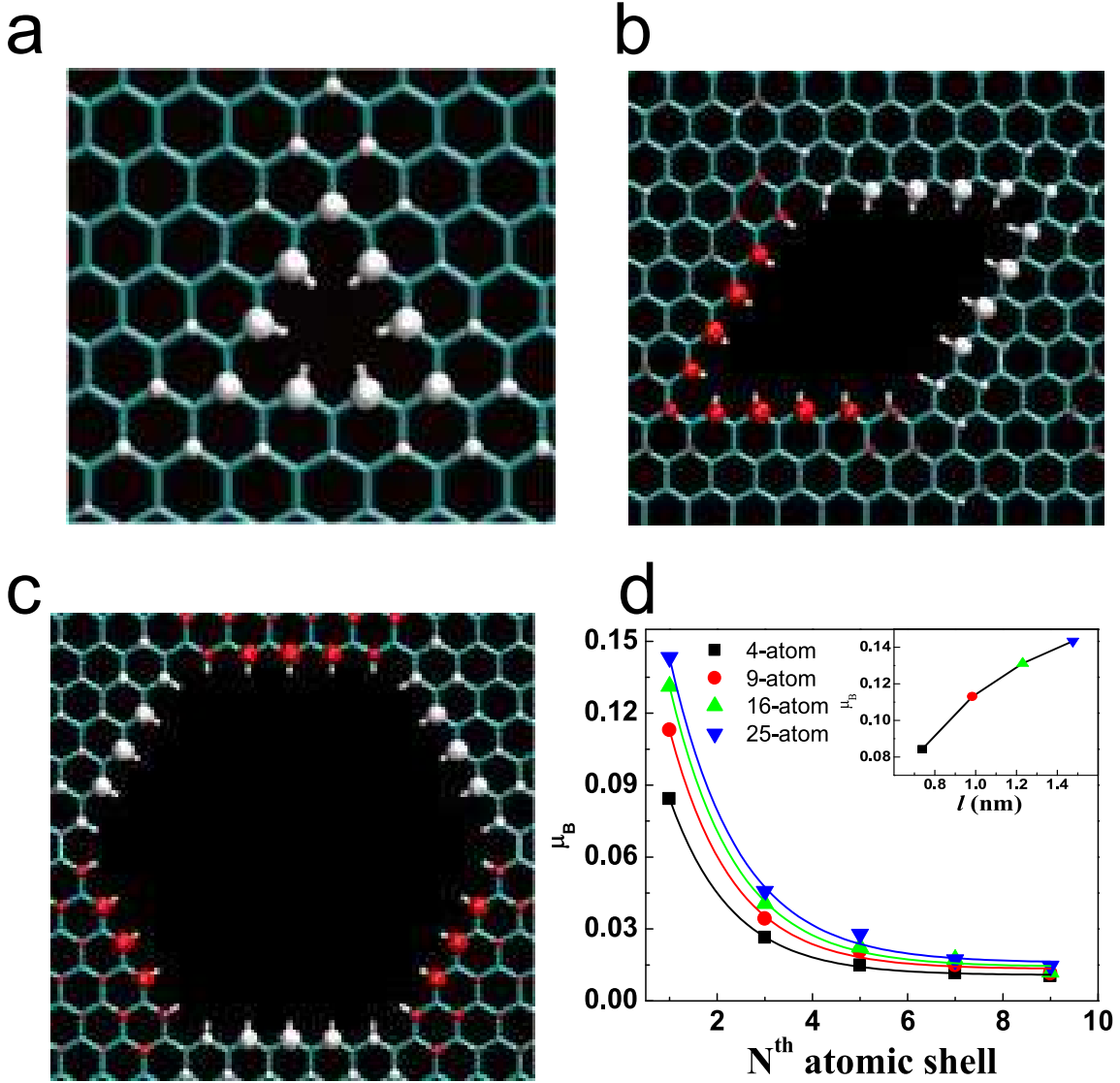}
\caption{\label{diffshape} The ground-state magnetic configurations
of different shapes of NHs. (a) FM triangular NH; (b) AF rhombus NH;
(c) AF hexagonal NH. In (a-c), white and red balls indicate the up-
and down-spin density isosurface at 0.02e/${\AA}$3, respectively;
blue and white sticks represent C-C and C-H bonds respectively.  (d)
The average local magnetic moment ($\mu_{B}$) per atom in the
triangular NH (Fig.\ref{diffshape}a) as a function of distance
moving away from the center of NH, measured in atomic shells with
the edge atoms as the first shell. The inset shows $\mu_{B}$ on the edge
vs. NH size ($l$). }
\end{figure}

Considering first a single zigzag NH by examining the
\emph{intra}-NH spin-spin interaction, we found that individual NH
can be viewed as an "inverse structure" of nanoflake or
nanoribbon, like an anti-flake or anti-ribbon, with similar spin
behavior. We determine the ground-state magnetism of three typical
NH shapes: triangular (Fig.\ref{diffshape}a), rhombus
(Fig.\ref{diffshape}b) and hexagonal (Fig.\ref{diffshape}c), by
comparing the relative stability of ferromagnetic (FM),
antiferromagnetic (AF) and paramagnetic (PM) configuration as a
function of NH size. Our calculations show that the ground state
is FM for triangular NHs, but AF for rhombus and hexagonal NHs,
and their corresponding spin densities are shown in
Fig.\ref{diffshape}a, \ref{diffshape}b and\ref{diffshape}c,
respectively. The magnetic moments are highly concentrated on the
edges and decay quickly away from the edge, as shown in
Fig.\ref{diffshape}d. Similar decaying behavior has been seen in
nanoribbons\cite{YWSon,Pisani} and
nanoflakes\cite{Fernandez,Wang}. The edge moment increases with
increasing NH size (see inset of Fig.\ref{diffshape}d).

The triangular NHs have a metastable ferrimagnetic state with two
edges having one spin and the other edge having the opposite spin
(see supplementary Fig.\ref{diffshape}). For a 4-atom triangular
NH (Fig.\ref{diffshape}a), the FM state is 52 meV lower in energy
than the ferrimagnetic state, and the latter is 13 meV lower than
the PM state. For a 32-atom rhombus NH (Fig.\ref{diffshape}b), the
AF state is 89.2 meV lower than the PM state; for a 54-atom
hexagonal NH (Fig.\ref{diffshape}c), it is 164.4 meV. The energy
difference increases with increasing NH size. The triangular NHs
favor FM at all sizes, whereas rhombus and hexagonal NHs only
become AF when the edge contains more than five atoms, i.e. they
are PM if the NH is too small. So, the triangular NHs have a
stronger tendency toward magnetization.

The magnetic ordering within a single NH is consistent with both the
theorem of itinerant magnetism in a bipartite lattice\cite{Lieb} and
the topological frustration model of the $\pi$-bonds\cite{Wang}
counting the unpaired spins in the nonbonding
states\cite{FLiu,Wang}. For a system like graphene consisting of two
atomic sublattices, each sublattice assumes one spin and the total
spin S of the ground state equals $\frac{1}{2}|N_{B}-N_{A}|$ where
N$_{B}$ (N$_{A}$) is the number of atoms on B (A) sublattice.
Because of the honeycomb lattice symmetry, atoms on the same zigzag
edge belong to the same sublattice; while atoms on two different
zigzag edges belong to the same sublattice if the two edges are at
an angle of 0$^{\textsf{o}}$ or 60$^{\textsf{o}}$, but different
sublattices if at an angle of 120$^{\textsf{o}}$ or
180$^{\textsf{o}}$. Consequently, the triangular NH are FM, because
all three edges are 60$^{\textsf{o}}$ to each other on the same
sublattice; the rhombus and hexagonal NHs are AF, because one-half
the edges are on the A-sublattice and another half on the
B-sublattice as the two types of edges are 120$^{\textsf{o}}$ to each other. This same argument can be applied to
nanoribbons\cite{YWSon} and nanoflakes\cite{Fernandez,Wang}.

Next, we consider GNH superlattices (a periodic array of NHs) by
examining the \emph{inter}-NH spin-spin interaction. In principle,
one can generate four out of five possible 2D Bravais lattices of
NHs (see supplementary Fig.\ref{LvsEandTc}). Here, we focus on the
honeycomb superlattices of triangular NHs (Fig.\ref{LvsEandTc}a and
\ref{LvsEandTc}b), in which each NH possesses a net moment acting
effectively as "one" spin. The superlattice contains two sublattices
of NHs, superimposed on the background of graphene containing two
sublattices of atoms. We realize that the NHs on the same sublattice
will be FM-coupled because their corresponding edges are at
0$^{\textsf{o}}$ to each other so that their edge atoms are on the
same atomic sublattice. On the other hand, the NHs on different
sublattices will be FM-coupled if they are in a parallel
configuration (Fig.\ref{LvsEandTc}a) but AF-coupled if they are in
an antiparallel configuration (Fig.\ref{LvsEandTc}b) when their
corresponding edges are at 180$^{\textsf{o}}$ to each other so that
their edge atoms are on different atomic sublattices. These have
indeed been confirmed by our first-principles calculations.
Independent of NH size and supercell dimension, the FM state is
favored for parallel configurations but the AF state is favored for
antiparallel configurations. In both cases, the spin-polarization
splits the edge states opening a gap at the Fermi
energy\cite{YWSon,Wang}. The total spin S in one unit cell equals to
$\frac{1}{2}|N_{B}-N_{A}|$; it increases linearly in the FM parallel
configuration but remains zero in the AF antiparallel configuration
with increasing NH size.

\begin{figure}
\includegraphics[width=6in]{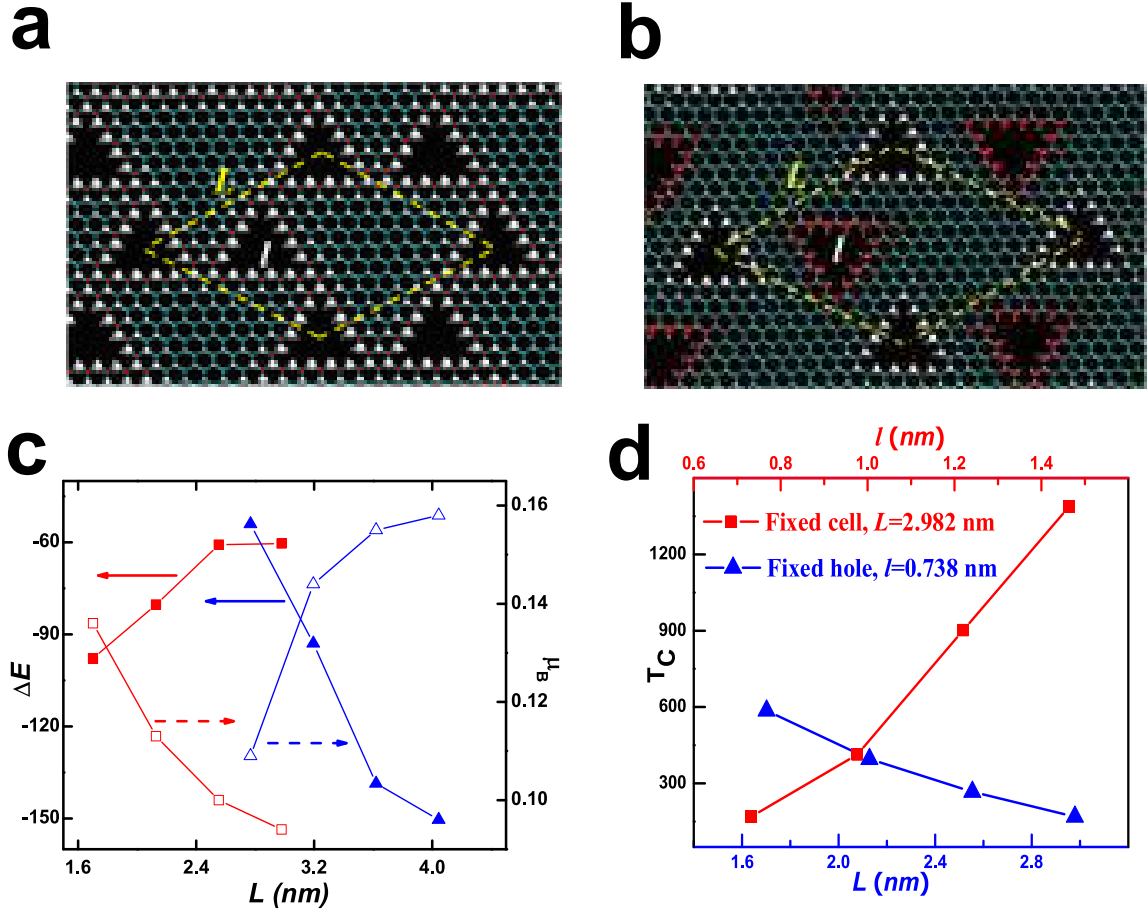}
\caption{\label{LvsEandTc} (a) Ground-state spin configurations in
a FM honeycomb NH superlattice. (b) Same as (a) in an AF
superlattice. All the symbols and notations for bonds and spin
densities are the same as Fig.\ref{diffshape}. Yellow dashed lines
mark the primitive cell. (c) $\Delta E_{pc}=E(FM)-E(PM)$ of the FM
superlattice (red solid squares) and $\Delta E_{ac}=E(AF)-E(PM)$
of the AF superlattice (blue solid triangles) versus cell
dimension ($L$); Edge magnetic moments, $\mu_{B}$ in the FM
lattice with fixed hole size ($l$=1.476 nm) (red open squares) and
in the AF lattice ($l$=0.738 nm) (blue open triangles)  versus
$L$. (d) Curie temperature of the FM superlattice as a function of
NH size ($l$) and $L$.}
\end{figure}

The collective magnetic behavior of a GNH superlattice depends on
inter-NH spin-spin interaction. Particularly, there exists super
exchange interaction between the NH spins, in addition to the spin
coupling defined by the underlying bipartite lattice (i.e., the
relative angle between the zigzag edges of GNHs). In
Fig.\ref{LvsEandTc}c, we plot  $\Delta E_{pc}=E(FM)-E(PM)$ for the
FM parallel configuration and $\Delta E_{ac}=E(AF)-E(PM)$ for the AF
antiparallel configuration as a function of cell dimension ($L$),
i.e., the NH-NH separation. $|\Delta E_{pc}|$ increases while
$|\Delta E_{ac}|$ decreases with decreasing $L$. This indicates that
as the NHs getting closer, the FM state becomes relatively more
stable than AF state in both configurations, i.e. the FM coupling is
favored by the super exchange interaction. Also plotted in
Fig.\ref{LvsEandTc}c are magnetic moments on the NH edges, which are
found to increase in the FM but decrease in the AF configuration
with decreasing $L$. This again reflects that the edge magnetization
on the neighboring NHs is enhanced with the same spin when they are
FM coupled but suppressed with the opposite spin when they are AF
coupled by the super exchange interaction.

The above results show that long-range ferromagnetic ordering can be
created by employing the parallel configuration of triangular NHs in
different lattice symmetries, as illustrated in supplementary
Fig. \ref{LvsEandTc}. The next important question is what Curie
temperature ($T_{c}$) they can have. We have estimated $T_{c}$ using
the mean-field theory of Heisenberg model\cite{Hynninen,Sato,Turek},
 \begin{equation}
 T_{c}=\frac{2\Delta}{3k_{B}}
 \end{equation}
Where $\Delta$ is the energy cost to flip one "NH spin" in the FM
lattice, which have been calculated directly from first principles
for the honeycomb lattices (Fig.\ref{LvsEandTc}a).
Figure \ref{LvsEandTc}d shows that $T_{c}$ increases from 169 K to
1388 K when NH size ($l$) increases from 0.738 to 1.476 nm with cell
dimension ($L$) fixed at 2.982 nm, and decreases from 586 K to 169 K
when $L$ increases from 1.704 nm to 2.982 nm with $l$ fixed at 0.738
nm. These trends are expected since magnetization is stronger for
larger NH size and higher NH density. Limited by computation time,
some of our cell dimensions are possibly unrealistically too small
(NH density too high), which gives rise to a very high $T_{c}$.
Still, it is important to point out that our calculations suggest
that it is possible to make FM GNH superlattices with $T_{c}$ above
room temperature by using a NH size of ~50 nm and a density of
$10^{-4}$ nm$^{-2}$, achievable by today's lithographic patterning
technology. We note that a recent experiment\cite{Barzola} has shown
a $T_{c}\geq$ 350 K in FM graphite made by proton bombardment.

\begin{figure}
\includegraphics[width=6in]{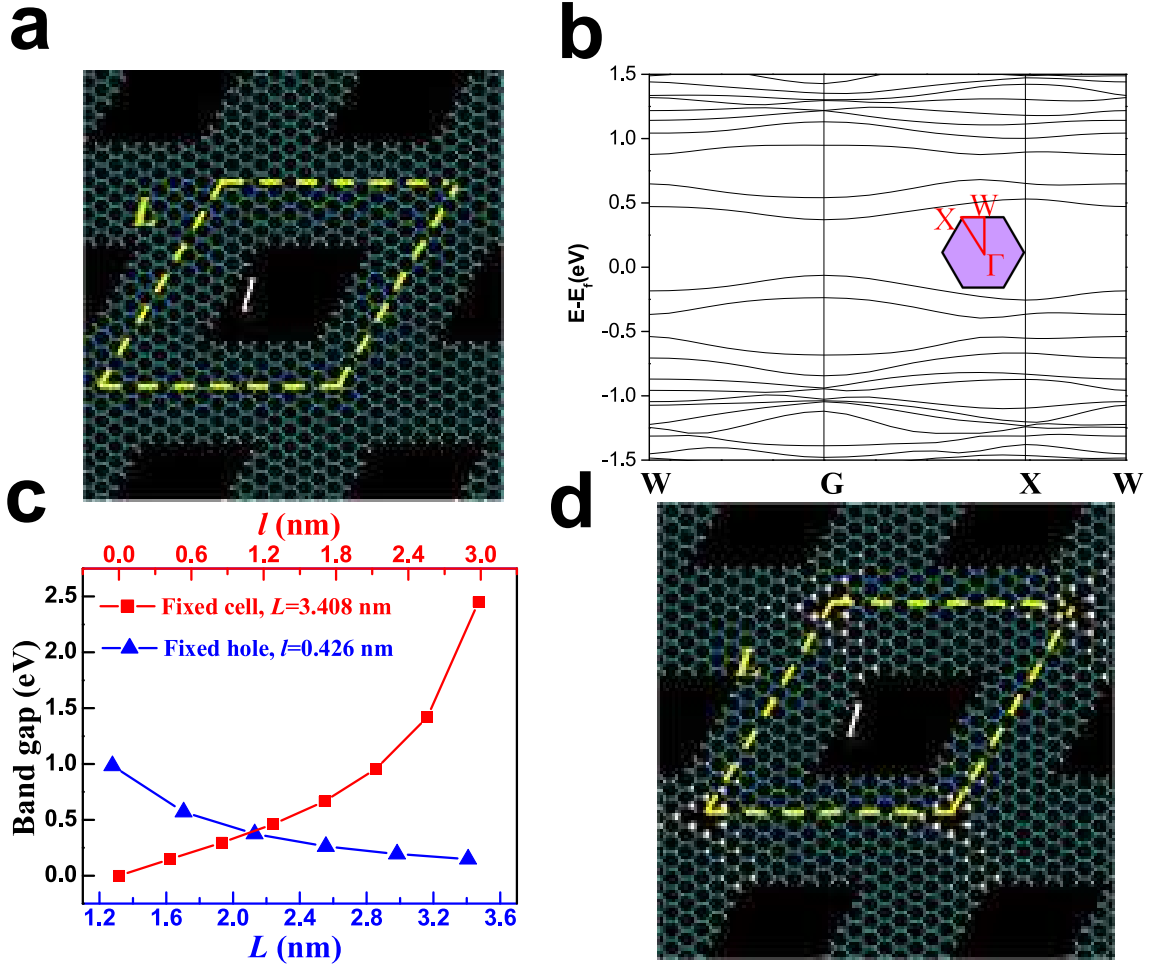}
\caption{\label{Mlattice} Illustration of DMS made from GNH
superlattice. (a) A semiconductor GNH hexagonal lattice
($L=8\sqrt{3}a$,$a=2.46\textrm{{\AA}}$ is the lattice constant of
graphene.) consisting of an array of rhombus armchair NHs
($L=8\sqrt{3}a$). (b) GGA Band structure of (a), the inset shows the
Brillouin zone. (c) TB band gap of (a)-type structures as a function
of NH size ($l$) and cell dimension ($L$). (d) Magnetic
semiconductor made by doped (a) with triangular zigzag NHs. All the
symbols and notations for bonds and spin densities are the same as
Fig.\ref{diffshape}.}
\end{figure}

Since graphene-based nanostructures hold great promise for future
electronics\cite{Novoselov1,Novoselov2,Ozyilmaz,Yan}, our discovery
of GNH magnetism offers the exciting prospect of combining magnetic
and semiconducting behavior in one material system. Here, we
demonstrate the possibility of making diluted magnetic
semiconductors (DMS) by exploiting GNHs with two different kinds of
edges. Similar to superlattices of zigzag NHs, we can create
superlattices of armchair NHs, which constitute a class of 2D
semiconductors. Figure \ref{Mlattice}b shows the band structure of a
superlattice of rhombus armchair NHs (Fig.\ref{Mlattice}a) having a
direct band gap of 0.43 eV, as obtained from first-principles
calculations. Figure \ref{Mlattice}c shows the band gap as a
function of NH size ($l$) and cell dimension ($L$), from
tight-binding calculations\cite{Our}. The gap increases with
increasing $l$ but decreases with increasing $L$.

DMS can be made by adding triangular zigzag NHs into the
semiconductor superlattice, as illustrated in Fig. \ref{Mlattice}d.
To ensure the ferromagnetism, all triangular NHs must be parallel
with each other acting like magnetic dopants. Usually DMS are
synthesized by mixing two different materials, typically III-V
semiconductors and transition-metal magnetic
elements\cite{Jungwirth,Macdonald}. The main challenge is to
increase the magnetic dopant concentration in order to raise the
Curie temperature, because the two types of materials are usually
not miscible. Here, we introduce an "all-carbon" DMS, in which
combined semiconductor and magnetic behavior are achieved by
structural manipulation. Consequently, room-temperature DMS are
possible because the dopant concentration can be increased without
the miscibility problem. One might also consider doping other
magnetic elements into the semiconducting GNH superlattice.

\begin{figure}
\includegraphics[width=6in]{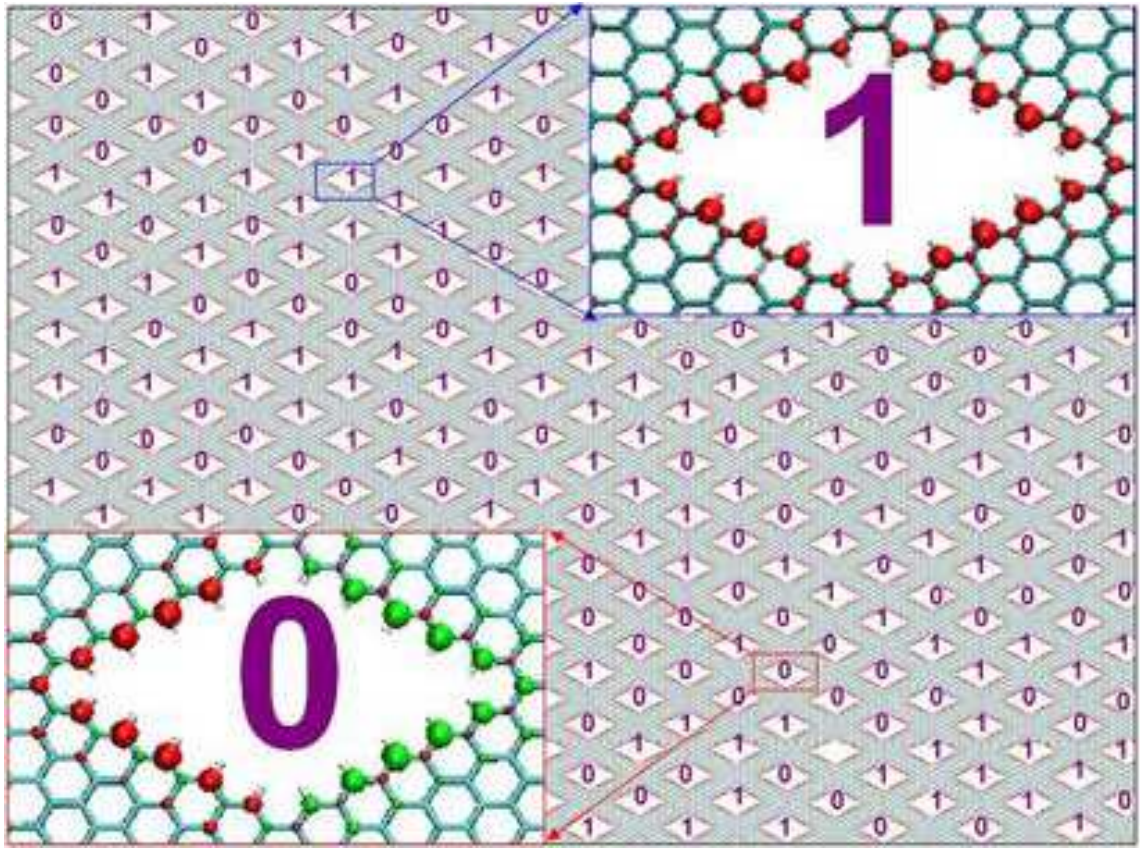}
\caption{\label{GNH} Schematic Illustration of a magnetic storage
medium consisting of a patterned array of rhombus GNHs. The insets
show the detailed structure of "0" and "1" bit, represented by the
ground-state AF configuration (S=0) and the excited FM configuration
(S=N), respectively. Red and green balls show the spin-up and
spin-down density at an isosurface value of 0.02$
e/\textrm{{\AA}}^{3}$.}
\end{figure}

It is very exciting to consider making engineered magnetic materials
with NHs for various applications. For example, it is possible to
directly pattern NHs into engineered magnetic storage media (see
Fig. \ref{GNH}). The ground state of rhombus NHs is AF (Fig.
\ref{diffshape}b and Fig.\ref{GNH}, lower-left inset) and their
first excited state is FM (Fig. \ref{GNH}, up-right inset) when the
NH size is larger than 14.6${\AA}$ according to our calculation.
Taking each NH as one bit, we can assign the ground state with "S=0"
and the excited state with "S=N" to represent the '0' and '1',
respectively. The switching between '0' to '1' can be done by
applying a local magnetic field or energy pulse to convert between
the ground and the excited state. Using a NH size of ~50 nm and a
density of 10$^{-4}$ nm$^{-2}$, a storage density about 0.1 terabit
per square inch would be achieved, much higher than the current
density in use. One interesting topic of future study is the
magnetocrystalline anisotropy around individual NHs, which must be
larger than $k_{B}T$ for the proposed storage media to work.

We thank DOE-NERSC and Center for High Performance Computing (CHPC)
at the University of Utah for providing the computing resources.
This work was supported by DOE.

\newpage

\begin{figure}
\includegraphics[width=3.5in]{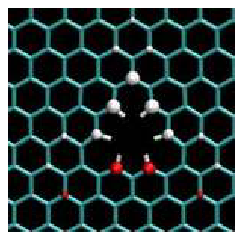}
\\Supplementary Fig.1 The spin-density plot of the ferrimagnetic
configuration of a 4-atom triangular NH. White and red balls
indicate the up- and down-spin density isosurface at
0.02e$/\textrm{{\AA}}^3$ respectively; blue and white sticks
represent C-C and C-H bonds respectively.
\end{figure}

\begin{figure}
\includegraphics[width=6in]{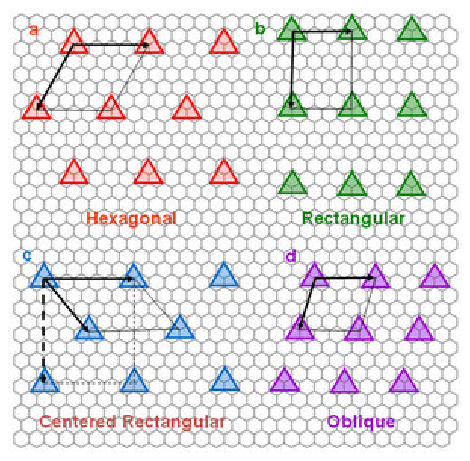}
\\Supplementary Fig.2 Schematic illustration of four possible types of
Bravais lattice of GNHs that can be patterned in graphene. Solid
arrows and lines mark the primitive cells.  (a) hexagonal lattice;
(b) rectangular lattice; (c) centered rectangular lattice; the
dashed lines mark the conventional cell; (d) oblique lattice. Note
that the square lattice is not possible.
\end{figure}


\begin{references}

\bibitem{Slater}
J. C. Slater,
\newblock Phys. Rev. {\bf 49}, 537 (1936).

\bibitem{FLiu}
F. Liu, S. N. Khanna, and P. Jena,
\newblock Physical Review B {\bf 42}, 976 (1990).

\bibitem{Shibayama}
Y. Shibayama, H. Sato, T. Enoki, et al.,
\newblock Physical Review Letters {\bf 84}, 1744 (2000).

\bibitem{Esquinazi}
P. Esquinazi, D. Spemann, R. Hohne, et al.,
\newblock Physical Review Letters {\bf 91}, 227201 (2003).

\bibitem{Lehtinen}
P. O. Lehtinen, A. S. Foster, Y. C. Ma, et al.,
\newblock Physical Review Letters {\bf 93}, 167202 (2004).

\bibitem{HLee}
H. Lee, Y. W. Son, N. Park, et al.,
\newblock Physical Review B {\bf 72}, 174431 (2005).

\bibitem{Coey}
J. M. D. Coey, M. Venkatesan, C. B. Fitzgerald, et al.,
\newblock Nature {\bf 420}, 156 (2002).

\bibitem{Kusakabe}
K. Kusakabe and M. Maruyama,
\newblock Physical Review B {\bf 67}, 092406 (2003).

\bibitem{YWSon}
Y. W. Son, M. L. Cohen, and S. G. Louie,  ,
\newblock Nature {\bf 444}, 347 (2006).

\bibitem{Nomura}
K. Nomura and A. H. MacDonald,
\newblock Physical Review Letters {\bf 96}, 256602 (2006).

\bibitem{Pisani}
L. Pisani, J. A. Chan, B. Montanari, et al.,
\newblock Physical Review B {\bf 75}, 064418 (2007).

\bibitem{Brey}
L. Brey, H. A. Fertig, and S. Das Sarma,
\newblock Physical Review Letters {\bf 99}, 116802 (2007).

\bibitem{Novoselov}
K. S. Novoselov, Z. Jiang, Y. Zhang, et al.,
\newblock Science {\bf 315}, 1379 (2007).

\bibitem{Jiang}
D. E. Jiang, B. G. Sumpter, and S. Dai,
\newblock Journal of Chemical Physics {\bf 127}, 124703 (2007).

\bibitem{Yazyev}
O. V. Yazyev and L. Helm,
\newblock Physical Review B {\bf 75}, 125408 (2007).

\bibitem{Fernandez}
J. Fernandez-Rossier and J. J. Palacios,
\newblock Physical Review Letters {\bf 99}, 177204 (2007).

\bibitem{Wang}
W. L. Wang, S. Meng, and E. Kairas,
\newblock J. Appl. Phys. Nano Letters{\bf 8}, 241 (2007).

\bibitem{Huang}
B. Huang, F. Liu, J. Wu, et al.,
\newblock \href{http://arxiv.org/abs/0708.1795v1.}{http://arxiv.org/abs/0708.1795v1.}

\bibitem{Kresse}
G. Kresse and J. Hafner,
\newblock Physical Review B {\bf 47}, 558 (1993).

\bibitem{Lieb}
E. H. Lieb,
\newblock Physical Review Letters {\bf 62}, 001201 (1989).

\bibitem{Hynninen}
T. Hynninen, H. Raebiger, and J. von Boehm,
\newblock Physical Review B {\bf 75}, 125208 (2007).

\bibitem{Sato}
K. Sato, P. H. Dederics, and H. Katayama-Yoshida,
\newblock Europhysics Letters {\bf 61}, 403 (2003).

\bibitem{Turek}
I. Turek, J. Kudrnovsky, G. Bihlmayer, et al.,
\newblock Journal of Physics-Condensed Matter {\bf 15}, 2771 (2003).

\bibitem{Barzola}
J. Barzola-Quiquia, P. Esquinazi, M. Rothermel, et al.,
\newblock Physical Review B {\bf 76}, 161403(R) (2007).

\bibitem{Novoselov1}
K. S. Novoselov, A. K. Geim, S. V. Morozov, et al.,
\newblock Science {\bf 306}, 666 (2004).

\bibitem{Novoselov2}
K. S. Novoselov, A. K. Geim, S. V. Morozov, et al.,
\newblock Nature {\bf 438}, 197 (2005).

\bibitem{Ozyilmaz}
B. Ozyilmaz, P. Jarillo-Herrero, D. Efetov, et al.,
\newblock Physical Review Letters {\bf 99}, 166804 (2007).

\bibitem{Yan}
Q. M. Yan, B. Huang, J. Yu, et al.,
\newblock Nano Letters {\bf 7}, 1469 (2007).

\bibitem{Our}
Our tight-binding band structure calculations for semiconductor
armchair GNH superlattices were performed using the
nearest-neighbor {$\pi$}-band model with the hopping parameter =
3.0 eV.

\bibitem{Jungwirth}
T. Jungwirth, J. Sinova, J. Masek, et al.,
\newblock Reviews of Modern Physics {\bf 78}, 809 (2006).

\bibitem{Macdonald}
A. H. Macdonald, P. Schiffer, and N. Samarth,
\newblock Nature Materials {\bf 4}, 195 (2005).



\end{references}
\end{document}